\documentclass{aastex}          %% The manuscript based on AASTeX v5.x
\usepackage{spr-astr-addons}    %% mimicing ASTR journal style
\begin{document}
%% Article title
%
%\title{A multi-wavelength approach to galaxy evolution: \\
%an ultraviolet investigation of  early type galaxies  in low density environments}

\title{A multi-wavelength study of the evolution of  Early-Type
Galaxies in Groups: the ultraviolet view}

%% Running heads
\shorttitle{A multi-wavelength study of Early-Type
Galaxies' evolution in Groups: the ultraviolet view}
\shortauthors{Rampazzo et al.}

%% Author and Affilations
\author{Rampazzo R.\altaffilmark{1}, Mazzei P.\altaffilmark{1}, Marino A.\altaffilmark{1}, Bianchi L.\altaffilmark{2}, Plana, H.\altaffilmark{3}, Trinchieri  G.\altaffilmark{4}, 
Uslenghi M.\altaffilmark{5}, Wolter A.\altaffilmark{4}} 
%\and 
%\author{ \altaffilmark{2}}
%\affil{}
\email{roberto.rampazzo@oapd.inaf.it} %% non-output

%% Alternate Affilations
\altaffiltext{1}{INAF-Osservatorio Astronomico di Padova, Vicolo dell'Osservatorio 5, 35122 Padova, Italy}
\altaffiltext{2}{The Johns Hopkins University, Dept. of Physics and Astronomy, 3400 N. Charles Street, Baltimore, Maryland 21218}
\altaffiltext{3}{Laborat\'orio de Astrof\'isica Te\'orica e Observational,
Universidade Estadual de Santa Cruz --
45650-000 Ilh\'eus - Bahia Brazil }
\altaffiltext{4}{INAF-Osservatorio Astronomico di Brera, Via Brera 28, 20121 Milano, Italy}
\altaffiltext{5}{INAF-IASF, via E. Bassini 15, I-20133 Milano, Italy}

%% Abstract
\begin{abstract}
The ultraviolet-optical color magnitude diagram of rich  galaxy groups is characterised by a
 well developed {\it Red Sequence}, a {\it Blue Cloud} and the so-called  {\it Green Valley}. 
Loose, less evolved groups of galaxies which are probably not virialized yet 
may lack a well defined Red Sequence.  This is actually explained in the
framework of galaxy evolution. 
We are focussing on understanding galaxy 
migration towards the Red Sequence, checking for signatures of such a transition
 in  their photometric and morphological properties.
We report on the ultraviolet properties of a sample of
early-type (ellipticals+S0s) galaxies inhabiting the Red Sequence.  
The analysis of their structures, as derived by fitting a 
S\'ersic law to their ultraviolet luminosity profiles, suggests 
the presence of an underlying disk. This is the hallmark of 
dissipation processes that still must have  a role in the 
evolution of this class of galaxies. Smooth Particle Hydrodynamic
simulations with chemo-photometric implementations 
able to match the global properties of our targets are used 
to derive their evolutionary paths through 
 ultraviolet-optical colour magnitude diagrams, providing some
fundamental information such as the crossing time through the
Green Valley, which depends on their luminosity. 
The transition from the Blue Cloud to the
Red Sequence takes several Gyrs, being about 3-5 Gyr for the
the brightest galaxies and more long for fainter ones, if it occurs.\\
The photometric study of nearby galaxy structures in ultraviolet is seriously 
hampered by either the limited field of view the cameras 
(e.g in {\it Hubble} Space Telescope) or by the low spatial resolution
of the images (e.g in the {\it Galaxy Evolution Explorer}). 
Current missions equipped with telescopes and 
cameras sensitive to ultraviolet wavelengths, such as
{\it Swift}-{\tt UVOT} and {\tt Astrosat-UVIT}, provide a
relatively large field of view and better resolution than the
 {\it Galaxy Evolution Explorer}.
 More powerful ultraviolet
instruments (size, resolution and field of view) are obviously bound 
to yield fundamental advances in 
the accuracy and depth of the surface photometry 
and in the characterisation of the galaxy environment.
% However,  in spite of the significant improvement in the accuracy
%and depth of the surface photometry and of the characterisation
%of the galaxy environment that the new ultraviolet missions
%will provide, they will still far short of the performances 
%obtained with optical surveys.     
\end{abstract}

%% Keywords
\keywords{Galaxies: elliptical and lenticular, cD --
 Galaxies: fundamental parameters --  ultraviolet: galaxies --
Galaxies: evolution-- interactions}

%%  Please use labels (\label, \ref) for section, figure, table, 
%%  equation  reference. Use \cite for bibliography references.
%
\section{Introduction}
\label{s:1}
The {\it Galaxy Evolution Explorer}, ({\tt GALEX} hereafter) 
\citep{Martin2005,Morrissey2007} opened an opportunity for direct mapping
of the star formation in galaxies in the 
Local Universe. Its large field of view (1.2$^\circ$ diameter)
enabled the first ultraviolet all sky survey 
\citep[e.g.][]{Gil2007,Bianchi2011}, which provided a new look at 
galaxy structures at these wavelengths, to be compared with the 
optical mapping \citep[see e.g.][and references therein]{Boselli14}.

Ultraviolet-optical color magnitude diagrams,  e.g. near ultraviolet
(NUV-$r$)  vs. M$_r$\footnote{$r$ is one of the Sloan Digital 
Sky Survey optical bands.},
help in defining an evolutionary scenario for galaxies.
Figure~1 of  \citet[][]{Salim07} shows this diagram for about 50.000
nearby ($z\lesssim0.1$), optically selected galaxies. Galaxies distributed
in two well populated  areas, a Red Sequence and a Blue Cloud 
separated by a relative large ($\simeq$ 3\,mag)  intermediate zone, the so-called 
Green Valley. The Blue Cloud is basically populated by late-type galaxies, 
while early-type galaxies  dominate the Red Sequence.
Since  \citet[][]{Salim07},  it became clear 
that  on-going and/or recent star formation is not restricted to the Blue Cloud. 
H$\alpha$ line emitters can be found in the Red Sequence  which
therefore includes galaxies with some star formation activity. 
{\it Bona-fide} early-type galaxies in the Green Valley, in particular 
S0s,  show H$\alpha$ nuclear emission more typical of 
 star forming galaxies. \citet[][]{Schawinski07} suggested that as many
as 30\% of early-type galaxies may have hosted recent star formation. 
However, it has been known for a long time that many, if not most, early-type
galaxies have low-ionization emission line regions (LINERs) 
\citep[see e.g.][and references therein]{Phillips86,Annibali10}, i.e.
their nuclear H$\alpha$ emission cannot be directly associated  
to star formation \citep[see e.g Figure~1 in][]{Salim07}  rather
to AGN emission.   

One of the {\tt GALEX} breakthroughs has been the discovery of 
extended ultraviolet emission in the galaxy outskirts \citep{Thilker07,
Thilker08}. Star formation has been detected in early-type galaxies
not only in their central part but also in the disk, or in rings/arm-like 
structures  \citep[see e.g.][and references therein]
{Rampazzo07,Marino09,Jeong09,Thilker10,Marino11a,Rampazzo11,Marino11b}. 

Here we review  our investigation on early-type galaxies in low density
environments, highlighting the importance of ultraviolet observations
which able to detect even extremely low star formation rates and
to constrain Smooth Particle Hydrodynamic (SPH hereafter) simulations with
chemo-photometric implementation  (SPH-CPI hereafter)
to determine galaxy evolution. In \S~2 we summarise 
results of our analysis of the (NUV-$r$) - M$_r$ color magnitude diagram
 of galaxy groups of different
richness and dynamical properties.  In \S~3 we discuss the results of our
multi-wavelength analysis of the luminosity profiles of 11 early-type galaxies observed with
{\it Swift}-{\tt UVOT}. In \S~4 we briefly describe the recipes of our SPH-CPI simulations 
able to reproduce the global properties of our targets, i.e. their absolute
magnitudes, spectral energy distribution (SED hereafter), morphology,  
and other galaxy properties that we will use to better understand 
their evolution. 
In \S~5 we summarize our results and preliminary conclusions, 
and derive  our expectations in  \S~6 from on-going and future 
ultraviolet missions. 

%% Figure 1
%
\begin{figure}%[tb]
\includegraphics[width=\columnwidth]{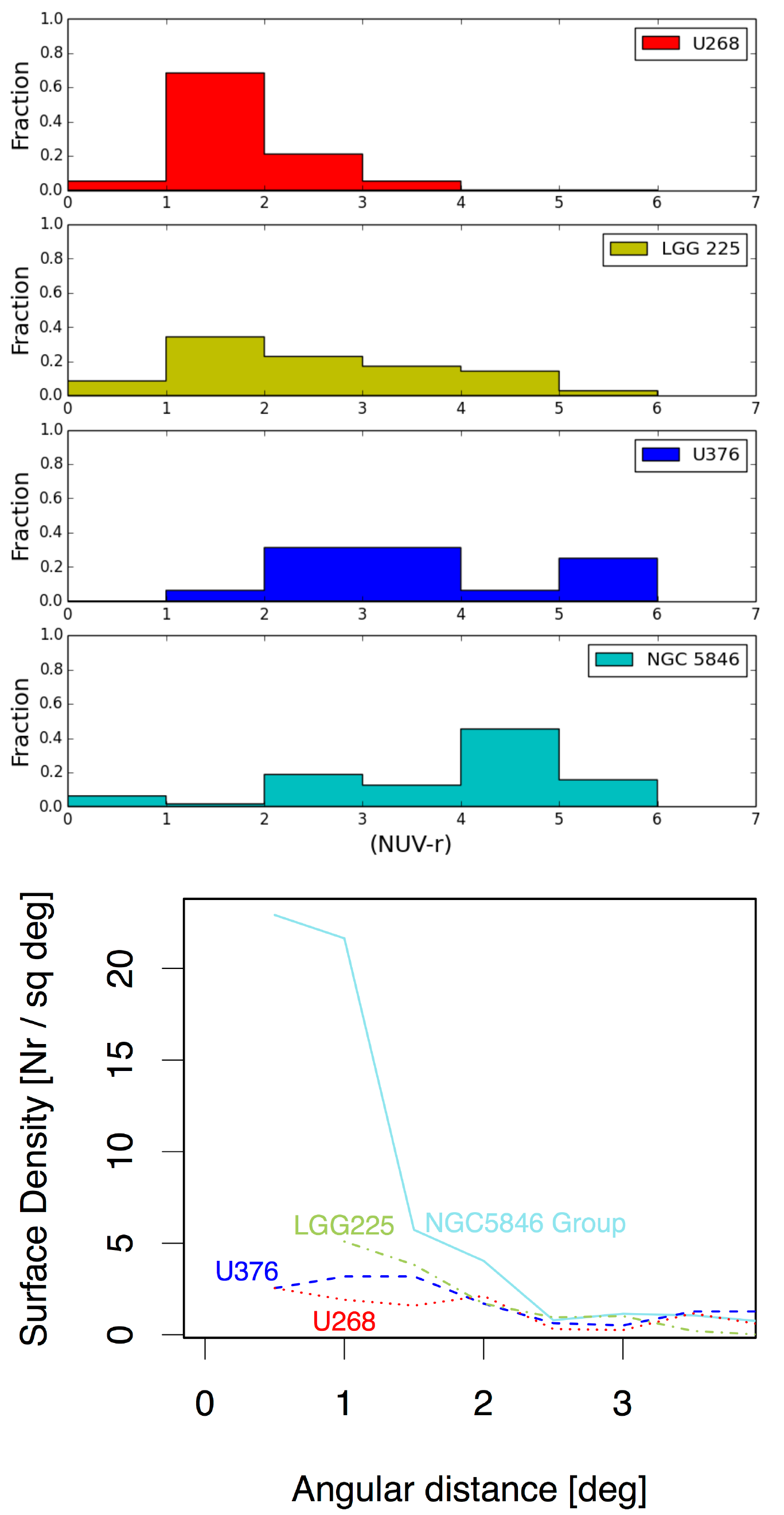}
\caption{(Top panel:)  The (NUV-r)  colour distribution  of galaxies  in four groups of increasing galaxy density, namely USGC U268, LGG 225, U376 and U677 (alias NGC 5846 group).
 Groups have been investigated in ultraviolet by \citet{Marino2010,Marino13, Marino16}. 
(Bottom panel) The surface galaxy density of  the above 
groups as a function of the angular distance from the group centre of mass. 
NGC 5846 is by far the densest group  with the richest Red Sequence. 
 The surface galaxy density of U268 is slightly above the galaxy background density
(adapted from \citet{Marino16})} 
%% no full stop at the end of caption
\label{fig:1}
\end{figure}
%%%%%%%%%%%%

\section{(NUV-$r$) vs. M$_r$ color magnitude diagram as a tracer of the co-evolution of galaxies 
in nearby groups}
\label{s:2}

Augustus Oemler and collaborators
remarked that ``the properties of many galaxies have evolved 
during recent epochs ... and  the end point of this evolution results 
in galaxy populations that vary over space ..'' 
\citep[see][and references therein]{Oemler17}. 
Groups are key structures to understand the path of this evolution
in low density environments and their
ultraviolet-optical  color magnitude diagrams are a useful 
tool to investigate the co-evolution of these cosmic structures 
and their member galaxies. We have obtained the
(NUV-$r$) - M$_r$ color magnitude diagram  of  four groups
USGC U268, LGG 225, U 376 and U677 (alias NGC 5486 group).
\citep[][and references therein]{Marino13,Mazzei14a,Marino16,Mazzei17}. 
These groups span a wide range both in galaxy populations and 
dynamical properties. Two of them, namely USGC U~268 and LGG~225 
are not yet virialized. These Local Group 
analogs are mostly populated by late-type galaxies. At odds,
the NGC~5846 group  is  the third largest nearby galaxy association
after Virgo and Fornax clusters. This group is dominated by early-type galaxies
and the dynamical analysis of its 99 known members indicates 
that it is virialized.

Assuming that the range 5$\lesssim(NUV-r)\lesssim$6.5 defines the 
Red Sequence region \citep[][]{Yi05,Salim07}, we derive the fraction 
of galaxies in this region in the above groups. The comparison between our virialized 
and not virialized groups is shown  in the top four panels of Figure~\ref{fig:1}. 
The galaxy population in the red sequence increases
from the loose, not virialized groups (USGC U268 and LGG~225) 
to the rich and virialized NGC 5846 group. 
Note also the increasing fraction of galaxies 
in the green valley, defined as 3.5$\lesssim(NUV-r)\lesssim$5,
from LGG~225 to USGC U376 and NGC 5846.
The bottom panel of Figure~\ref{fig:1} shows  the group surface density
as a function of distance from the group centre of mass. 
The surface density varies greatly from USGC~268 to NGC 5846 
and so does the galaxy population, from late-type galaxies  in loose
groups, to early-type in more dense ones. In the reasonable assumption
that these are all aspects of different phases of group evolution,
less evolved, more primitive groups have indeed a shallower surface density
than the virialized NGC 5846 group. Groups and galaxies seem
 indeed co-evolving via different mechanisms, e.g. via encounters and merging, 
whose effects still need to be deeply investigated in terms of
 morphological, structural and star formation 
consequences \citep[see e.g.][and references therein]{Weigel17}.  

One question raised by the analysis of group color magnitude diagrams 
is related to the morphology of galaxies in the red sequence and in the 
green valley. Are the galaxies in the Red Sequence, mostly ellipticals and S0s,
fairly unperturbed, at odds with galaxies in the Green Valley?
Far (FUV) and near (NUV) ultraviolet images of galaxies 
in our groups obtained by {\tt GALEX} show that
signatures of interaction are rare  both in loose and  dense groups. 
For example, 1 out of 4 members in LGG~225  either shows 
morphological distortions or are considered pairs
\citep{Mazzei17} while NGC~5846 has a much lower fraction.
However, galaxy-galaxy encounters are  the main drivers of the evolution
within groups. An intriguing case is that of the pair 
 NGC 3447/3447A in LGG~225,  highlighting  the problem of ``false'' galaxy pairs
and the  growth of disk instabilities in galaxy groups
we recently analysed in \citet[see\S~\ref{s:4.2}]{Mazzei17}. 
Mergers,  which can be  the final phase of an encounter,
alter  the galaxy morphology  and may fuel  the 
star formation rate via proper gas pipelines. We have considered
the interesting case of the  NGC~454 system, 
a galaxy pair which is still in the merging process   \citep{Plana17}.
Signatures of past interactions are given by
asymmetries  or morphological distortions. However,
some structures, such as bars and rings, can also arise
(therefore, we pay attention also to these features). 
In particular, rings are  enhanced in ultraviolet and far from 
being as symmetric as in the optical bands \citep{Rampazzo17}.

%% Figure 2
%
\begin{figure*}[tb]
\center
\includegraphics[width=12cm]{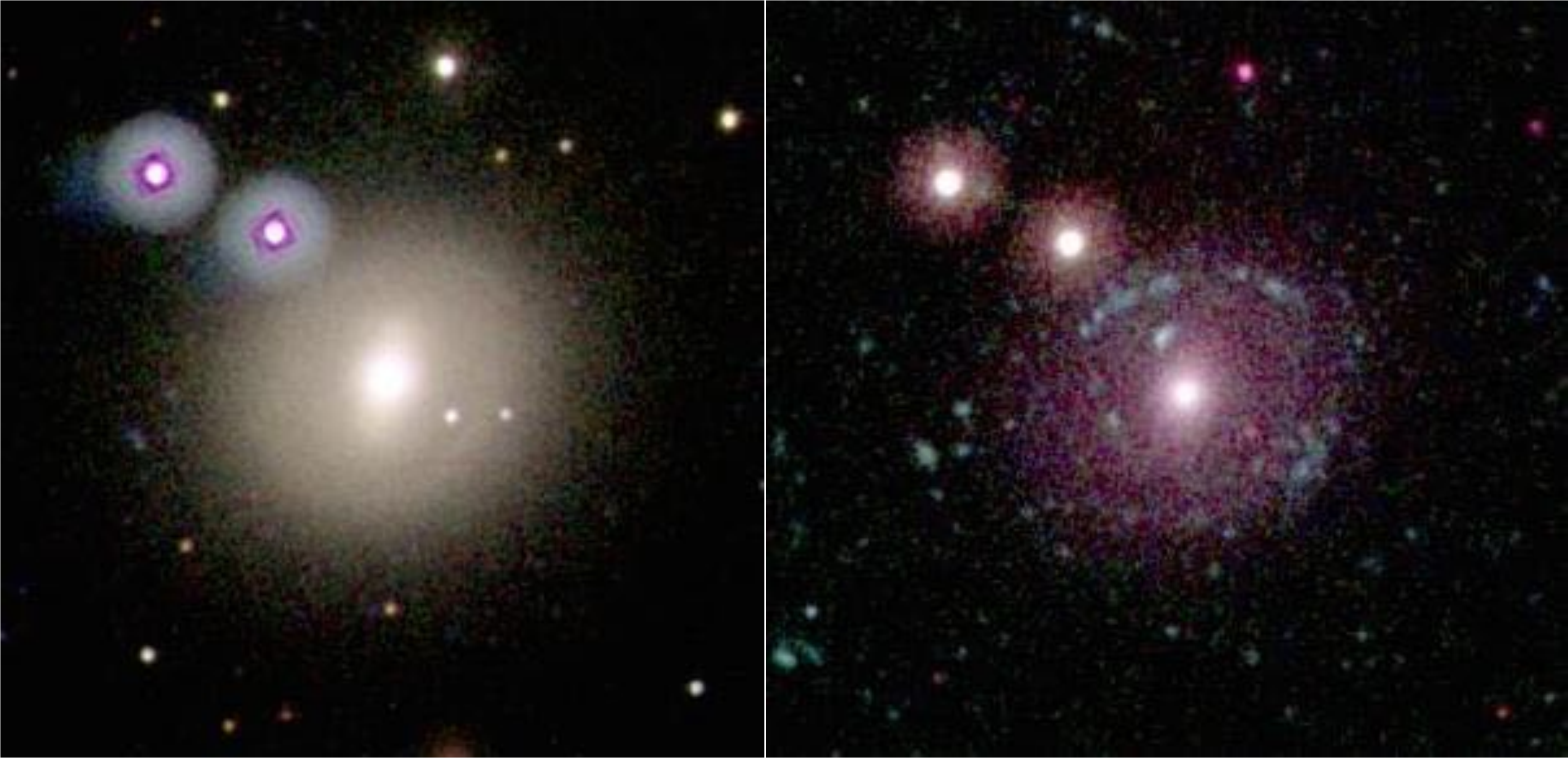}
\caption{{\it Swift}-{\tt UVOT} images of NGC~1533 in the Dorado group. 
{\it Left panel}: colour composite image in the U, B, V filters 
(U=blue, B=green, V=red) and, {\it right panel}, in the  W1, M2 and W2 filters 
(W2=blue, M2=green, W1=red). 
The field of view is  5\arcmin$\times$5\arcmin,  
North is on the top, East to the left  \citep{Rampazzo17}. Bright ring/arm-like structures are
detected in ultraviolet.  Furthermore, some of the ultraviolet bright regions, visible
in the South-East region of the field, likely belong to NGC 1533. Indeed, the galaxy extends 
 far beyond the optical outer ring and it is embedded in a huge HI envelope connecting it
 to IC 2038 \citep[see][and references therein]{Werk10}} 

%% no full stop at the end of caption
\label{fig:1b}
\end{figure*}
%%%%%%%%%%%%

To go further with our analysis  we selected 
groups which may mark  the transition from the
non virialized to the virialized phase. These intermediate 
phases are expected to be characterized by  sub-structures
and clumps in groups populated by some early-type galaxies
with interaction signature such as shells and/or rings. 
Early-type galaxies should represent
the end-product of galaxy evolution: mapping their ultraviolet-optical properties
in particular when they are the bright end of a group population,
gives precious  insight of the main evolutionary mechanisms 
at work in the group itself.

%--------------------------------------- Table 1 ----------------------------------------------
\begin{deluxetable}{lrrrrrrr}
\tabletypesize{\scriptsize}
\rotate
\tablecaption{Global properties of the \cite{Rampazzo17} sample \label{tbl-1}}
\tablewidth{0pt}
\tablehead{
\colhead{Galaxy} & \colhead{D$_{25}$} & \colhead{D} & \colhead{scale} & \colhead{m-M} & \colhead{M$_{B}$ } & \colhead{M$_{HI}$} & \colhead{L$_{X}$(gas)} \\
\colhead{Ident.} & \colhead{[arcmin]} & \colhead{[Mpc]} & \colhead{[kpc\,arcmin$^{-1}$]} & \colhead{[mag]} &
\colhead{[mag]} & \colhead{[10$^9$\,M$_{\odot}$]} & \colhead{[10$^{40}$\,erg\,s$^{-1}$]}}
\startdata
   NGC 1366&2.1 & 21.1$\pm$2.1 & 6.1&  31.62$\pm$0.50 & -18.88$\pm$0.54 & $<$1.0 & $<$0.03\\
   NGC 1415&3.7 & 22.7$\pm$2.5 & 6.5&  31.78$\pm$0.55 & -19.23$\pm$0.59 & 1.2$^a$ & 0.1    \\
   NGC 1426&2.9 & 24.1$\pm$2.4  & 7.0  & 31.91$\pm$0.50 & -19.70$\pm$0.52 & ....   & $<$0.03 \\
   NGC 1533 &3.2& 21.4$\pm$2.1 & 6.2  &  31.65$\pm$0.50 & -19.86$\pm$0.52   &  7.4$^b$   & $<$0.11\\
   NGC 1543 &3.6& 20.0$\pm$2.0 & 5.8 & 31.50$\pm$0.50 & -20.11$\pm$0.53  & 0.8   &   $<$0.16 \\
   NGC 2685 &4.4& 16.0$\pm$1.6  & 4.8 &  31.02$\pm$0.50& -19.09$\pm$0.51  &  3.0$^c$  & $<0.04$ \\
   NGC 2974 &3.5& 21.5$\pm$2.0  & 6.2 &  31.66$\pm$0.46&-20.01$\pm$0.48 & 0.7$^d$   &  0.2  \\
   NGC 3818 &2.4& 36.3$\pm$3.6 & 10.4  &  32.80$\pm$0.50 & -20.22$\pm$0.58  & ...  & 0.55\\
   NGC 3962 &4.2& 35.3$\pm$3.5  & 10.2 & 32.74$\pm$0.50 &   -21.29$\pm$0.53 & 2.8$^e$   & 0.33\\ 
   NGC 7192 & 2.4&37.8$\pm$3.8  & 10.7 &  32.89$\pm$0.50 & -20.81$\pm$0.51   & 0.7$^e$ & 1.0\\
   IC 2006  &2.3 &20.2$\pm$2.0 & 5.9 &  31.53$\pm$0.50 & -19.34$\pm$0.51 & 0.3 & 0.08\\   
\enddata
\tablecomments{The apparent diameters (col. 2)  and the  adopted distances 
(col. 3) are derived from the Extragalactic Distance Database 
(EDD: http://edd.ifa.hawaii.edu), as in Papers~I and II.
Absolute total magnitudes in col. 6 are derived from col. 5 
using B-band observed total magnitudes and
extinction corrections from {\tt Hyperleda} \citep{Makarov2014} catalogue.
The HI masses (col. 7) are obtained using the distance in col. 3 and fluxes  
from {\tt NED} and from the following references: 
 $^a$  \citet{Courtois15}; $^b$ \citet{Ryan-Weber2003}; $^c$\citet{Jozsa09}; $^d$ \citet{Kim88}; $^e$ \citet{Serra10}.
X-ray gas luminosity  (col. 8) is from Table 7 of \citet{Trinchieri15}. The table is
adapted from \citet{Mazzei18}}
\end{deluxetable}
%--------------------------------- end Table 1 ---------------------------------------------------

To this end, we have selected the nearby, very extended (several square degrees)
Dorado group and its numerous and diversified galaxy population \citep[see][]{Firth06} for 
imaging observations at high resolution  with the OmegaCam@VST 2.6m ESO telescope  
at Cerro Paranal (Chile).  
Dorado is an example of a still evolving group, as suggested by 
its clumpy  structure,  the significant number of 
on-going and past interaction signatures, suc as the shell structure
of NGC 1549 and NGC 1553 \citep{Malin83},  and the presence 
of  kinematic anomalies \citep{Rampazzo88}. 
We have already investigated the peculiar ultraviolet structure of NGC 1533 in Dorado
with {\tt GALEX} and studied it in the context of galaxy evolution with SPH-CPI simulations
 \citep[see e.g.][]{Marino11a,Mazzei14b}.
 Figure~\ref{fig:1b} exemplifies our findings
by comparing optical and ultraviolet colour composite images. \\
%We  emphasise the information coming from
% the photometric structure of this galaxy
%and other bright ETGs in low density environments using {\it Swift}-{\tt UVOT}
%in the following section. 

{\it Swift}  offers a new perspective to study galaxies \citep{swift1,swift2,swift3}.
It is equipped with the 30cm {\tt UVOT} telescope 
with a relatively large FoV (17\arcmin$\times$17\arcmin), 
$W2$ ($\lambda_0 \sim 2030$\AA), $M2$ ($\lambda_0 \sim 2231$\AA), 
$W1$ ($\lambda_0 \sim 2634$\AA) ultraviolet filters and a PSF 
(FWHM=2\farcs92 for $W2$, 2\farcs45 for $M2$,
2\farcs37 for $W1$) significantly improved with respect to {\tt GALEX}. 
This PSF is, in general, still to large to study 
the bulge of nearby galaxies. Therefore, {\it Swift}-{\tt UVOT} data  
are useful to analyse the main body and the 
galaxy outskirts, which have revealed unexpected features
 useful to understand the evolutionary history of galaxies
\citep{Rampazzo17}.

%% Figure 3
%
\begin{figure}%[tb]
\includegraphics[width=\columnwidth]{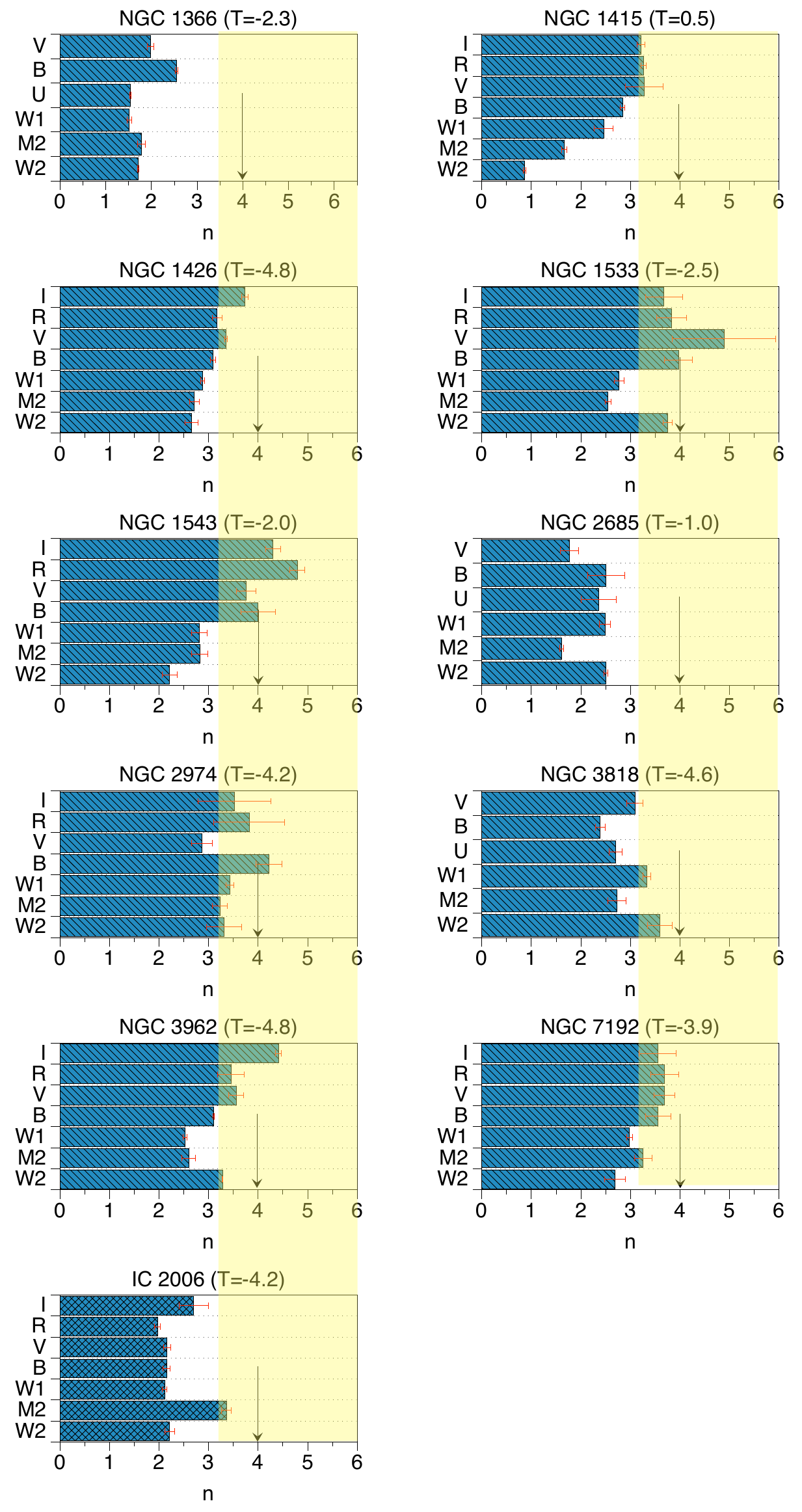}
\caption{Synoptic view of the Sersic index,  $n$, 
 obtained best fitting galaxy luminosity profiles  from near infrared
to near ultraviolet. The Sersic law \citep{Sersic68} is a generalisation of the de Vaucouleurs
r$^{1/4}$ law \citep{deVauc48}. The  value $n=1$ traces the presence
of an exponential disk, $n=0.5$ indicates a Gaussian shape.  The near
ultraviolet (W2, M2, W1) $n$ values, in the range 2$\leq n\leq$3,  tend to be lower 
than optical (B, V, R) and near infrared (I) ones and suggest the presence
of an underlying disk.
The arrow indicates $n=4$, the classical de Vaucouleurs law;   the
shadow highlights  values larger than 3. Galaxy identifier and
 morphological type, T,  where 5 $\leq T \leq -3$ is for Es and -3$< T \leq 0$
for S0s,  is given on the top of each panel \citep[figure adapted from][]{Rampazzo17}
} %% no full stop at the end of caption
\label{fig:3a}
\end{figure}
%%%%%%%%%%%%

\section{The ultraviolet surface brightness structure with {\it Swift}
and the evolution of early-type galaxies }
\label{s:3}

We start our multi-wavelength investigation of bright early-type galaxies 
in low density environments analyzing  11 galaxies \citep{Trinchieri15, Rampazzo17} 
using {\it Swift} (see Table~\ref{tbl-1}). 
All these galaxies are located either in groups or at the edge
of  rich associations, like Fornax,  where  galaxy densities 
are comparable to those of groups.

These galaxies have been selected from the Revised Shapley Ames 
catalogue \citep{RSA} and investigated in mid infrared 
by \citet{Rampazzo13} using the {\it Spitzer}-{\tt IRS} spectrograph.
This study 
identified galaxies with clear signatures of AGN 
and/or  on-going star formation activity in their nuclear region 
($\lesssim$3 $r_e/8$ ). Our aim is to look for residual recent 
star formation in the galaxy outskirts which will appear prominent in ultraviolet,   
and we used mid infrared nuclear spectra to select early-type  
galaxies that are either passively evolving or with a low degree
of activity.

%{\it Swift}-{\tt UVOT} images have been taken with the ``filter of the
%day" mode during {\tt XRT} observations and have been registered and co-added.
In ultraviolet images about 50\% of early-type galaxies 
show the presence of bright ring/arm-like
structures \citep{Rampazzo17}. We investigated the overall structure by fitting  
the ultraviolet, optical and near infrared surface brightness 
profiles  with a single S\'ersic law \citep{Sersic68}, 
deriving the  S\'ersic index, $n$, for each wavelength band.
Figure~\ref{fig:3a} shows the synoptic view of S\'ersic indices
obtained by fitting the same galaxy region.
Ultraviolet indices are generally  lower than optical 
and near infrared ones,  in the range $n \sim 2- 3$, suggesting the presence
of an underlying disk structure prominent at these wavelengths. 
Similar analyses, limited to optical and near infrared bands, have been performed by 
\citet{LaBarbera10}, \citet{Vulcani14}, and \citet{Kennedy16}.  These studies 
outlined  that the S\'ersic indices increase with  wavelength for 
late-type galaxies while they are almost  independent of wavelength  for 
early-type ones. The S\'ersic index  behaviour for late-type galaxies is
interpreted with the presence of a disk,  due to a radial  change in the
galaxy stellar populations and/or dust reddening.
With the extension of our investigation to  the ultraviolet range, we were able
to discover a similar trend in the early-type population which allowed us to suggest
for these ``red and dead'' galaxies the presence of an underlying 
disk \citep{Rampazzo17}. This  presence indicates that dissipative
processes are still working in these early-type galaxies. 

Multi-wavelength observations  (Table~\ref{tbl-1}) give important
constraints  on simulations. In the following section we present the
recipes of our SPH-CPI simulations, focussing on results for a few
cases. The complete analysis of each galaxy and  
results from the entire sample are still ongoing \citep{Mazzei18}.

\section{Modelling galaxy formation and
 evolution}
\label{s:4}

We use a grid of Smooth Particle Hydrodynamics simulations 
with Chemo-Photometric implementation (SPH-CPI hereafter)
to investigate the evolution of galaxies
\citep{Mazzei14a, Mazzei14b}. 
All simulations  start from collapsing triaxial
systems composed of dark matter  and gas, as in \citet[][]{Mazzei03}.
The  simulated  halos, as detected by the Hubble flow, share all the same
 initial conditions: the virial ratio (0.1),  the average density, and the spin parameter.
In more detail, each system is built up with a spin parameter, 
$\lambda$, given by $|{\bf J}||$E$^{0.5}$/(GM$^{0.5}$), where E
is the total energy, J is the total angular momentum, and G is the 
gravitational constant; $\lambda$ is equal to 0.06 and aligned to the 
shorter principal axis of the dark matter halo. 
The initial triaxiality ratio of the dark matter halos, 
$\tau$ = (a$^2$ - b$^2$)/(a$^2$ - c$^2$),  is 0.84 
\citep[][their Table 1]{Mazzei14a, Mazzei14b, Schneider12, CMM06} where a $>$ b$>$c.
The simulations include self-gravity of gas, stars and dark matter, radiative cooling, 
hydrodynamical pressure, shock heating, viscosity, star formation,
 feedback from evolving stars and type II supernovae, and chemical enrichment.
A Salpeter initial mass function (IMF) (0.01 - 100 M$_\odot$) is adopted, 
although alternative  IMFs have been tested \citep{Mazzei03}. The chemo-photometric
implementation is based on Padova evolutionary population synthesis  
models including stellar populations of six metallicity values sZ=0.0004, 0.001,0.004,0.008,
0.02,0.05.  The SPH-CPI simulations provide the spectral energy distribution
(SED) from 0.05 $\mu$m to 1 mm at each snapshot ($\Delta t$=3.7$\times$10$^7$ years), 
accounting for extinction and re-emission by dust  in a self-consistent way
 \citep[see][for details and references]{Mazzei14a,Mazzei14b,Mazzei17}.
 All the model parameters have been tested in previous works
 \citep[][and references therein]{Mazzei03} devoted to analysing the evolution 
 of the global properties of isolated triaxial system initially composed of dark matter and gas.

%% Figure 4
%
\begin{figure}[tb]
\includegraphics[width=\columnwidth]{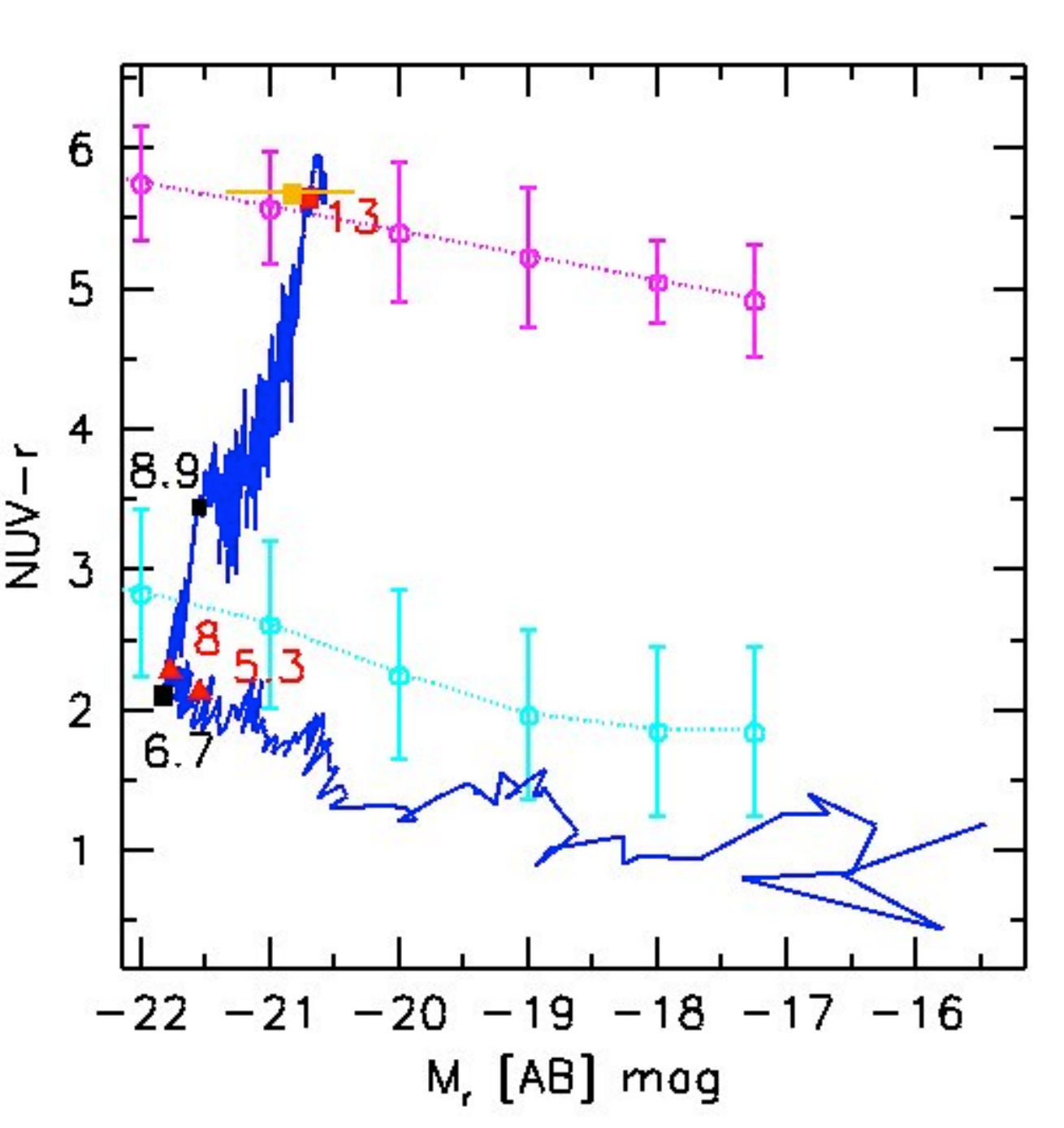}
\caption{ The evolution of NGC 1426 in the (NUV - $r$) - M$_r$ color magnitude 
diagram from SPH-CPI simulations. 
The Red Sequence  and the Blue Cloud  are indicated with magenta and cyan 
dotted lines, respectively. The open dots along the two sequences and the
error bars represent respectively the average value and the dispersion calculated
in bins of magnitudes from the \citet{Wyder2007} observations. 
The blue solid line shows the  evolutionary path of NGC 1426 
in the rest-frame, according to our SPH-CPI simulation matching the global 
properties of this galaxy at $z=0$. The numbers are ages in Gyr corresponding to remarkable 
phases in the evolution of the galaxy } 
%% no full stop at the end of caption
\label{fig:3}
\end{figure}

%%%%%%%%%%%%%
%%       Figure 5
%
\begin{figure*}%[tb]
\center
\includegraphics[width=12cm]{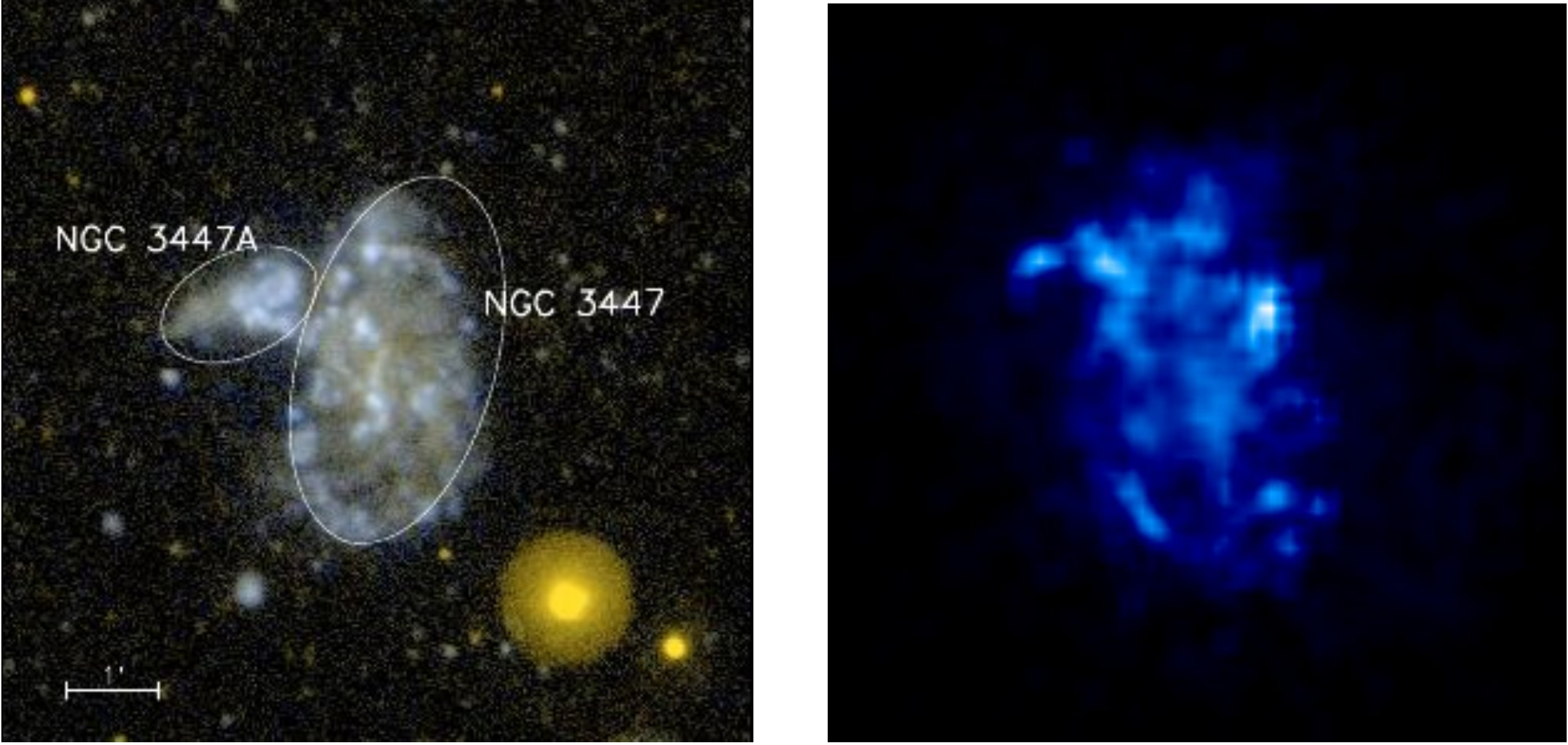}
\caption{ Comparison between the {\tt GALEX} NUV (yellow), FUV (blue) 
composite image, 7\arcmin$\times$7\arcmin, of \citet{Marino2010} {\bf(left panel)}, 
and the FUV band luminosity density map of the snapshot  best fitting the 
global properties of NGC~3447 normalised to the total flux in the box  
{\bf (right panel)}; maps are on the same spatial scale, and resolution (5\arcsec).
The mass of stars in this region is 2.2$\times$10$^9$\,M$_{\odot}$, that of the 
dark matter 13.7 times larger. Maps are on the same scale as observations, 
 in particular 7\arcmin$\times$7\arcmin  correspond to  a  projected 
 box  of 40\,kpc$\times$40\,kpc  on the stellar mass centre,
accounting for a scale of 5.5 kpc/\arcmin 
with H$_0$=75\,km\,s$^{-1}$\,Mpc$^{-1}$\citep{Marino2010} and the cosmological 
parameters:  $\Omega_{\Lambda}$=0.73, and $\Omega_{bar}$=0.27
(adapted from \citet{Mazzei17})
%given by the {\tt NED} catalogue.
} 
 \label{fig:4}
\end{figure*}
%%%%%%%%%%%%% END Figure 4/NGC3447

We performed a grid of simulations of mergers and encounters
between galaxies with different mass ratio (1:1 $\div$ 10:1) and gas fraction, 
exploring different orbital parameters,
starting from systems built up with the same initial
conditions and using as model parameters those tuned in the above
cited papers. For each early-type galaxy under study, a large set 
of SPH-CPI simulations is explored until
its absolute magnitude (B-band for example), multi-wavelenght SED and morphology
are matched.  The galaxy kinematics, when available,
is  used to further  constrain the simulation \citep[e.g.][]{Mazzei14b, Mazzei17}.
We are investigating all the galaxies in Table~\ref{tbl-1}  \citep[][in preparation]{Mazzei18}.
In the next two sub-sections  we focus on two examples of the above set:
NGC 1426, showing no peculiar features,   and NGC 1533 showing 
bright ring/arm-like structure in ultraviolet (in \S~\ref{s:4.1}).  In \S~\ref{s:4.2} we present 
two remarkable examples of the evolution of early and late-type galaxies in low density
environments.

\subsection{Early-type galaxies in groups}
\label{s:4.1}
 
NGC 1426 is classified as an E4 but our multi-wavelength structural
 analysis shows that there is a disk, 
suggesting a new classification as S0.  The galaxy is located in the Fornax  
Eridanus cloud with an environmental galaxy density
of $\rho=0.66$ galaxy Mpc$^{-3}$ \citep{Tully88}. 
No peculiar features, neither in the near nor in the far ultraviolet, are revealed
by both {\it GALEX} and {\it Swift} observations. NGC~1426 is considered 
an old and passively evolving galaxy. Using Lick line-strength indices
\cite{Annibali07} calculated that the galaxy has
an optical luminosity weighted age of 9.0$\pm$2.5 Gyr 
and nearly solar metallicity, $Z=0.024\pm0.05$. 
Mid infrared  spectroscopic  observations of its nuclear region  
\citep[][MIR class=0]{Panuzzo11, Rampazzo13} indicate that
there are neither emission lines \citep{Annibali10}, nor Polycyclic
Aromatic Hydrocarbon emissions,  signatures of a 
recent star formation episode \citep{Vega10}. NGC~1426 is not 
detected in HI, and \citet{Trinchieri15} measured 
an upper limit to the X-ray luminosity of its hot gas of 
 log L$_X<$38.95 erg~sec$^{-1}$.
 
Figure~\ref{fig:3} shows the evolutionary path,
 in the rest-frame (NUV- $r$) - M$_{r}$  color magnitude diagram,
predicted for NGC~1426 by the SPH-CPI simulation which
best fits its global properties. The derived SED, the morphology and
total magnitude, including the ultraviolet and optical luminosity 
profiles, are consistent with all the observational 
constraints \citep[][in preparation]{Mazzei18}. 
In particular, according to this simulation, the total B-band absolute 
magnitude of the galaxy is  -19.4$\pm$0.5 vs. -19.7$\pm$0.5\,mag 
as observed (Table~\ref{tbl-1}).  

 The evolution of NGC 1426 is driven by a major merger  between two 
halos of equal mass (M$_{tot}=2\times 10^{12}$ M$_\odot$).
 At the beginning the simulation considers 8$\times$10$^4$ particles to describe the halos 
of gas and dark matter. The halos centers of mass are moving,  at a separation of 654\,kpc, 
 with a relative velocity  of 142\,km~s$^{-1}$.
The red triangles in Figure~\ref{fig:3} mark the position of the galaxy  at 
$z=0.5$,  corresponding to a galaxy age of  8\,Gyr,  and at $z=1$, to 5.3\,Gyr.  
The simulation shows that the star formation rate drops significantly when 
the galaxy is older than  6.7\,Gyr, indicated with the  big black square in the figure.
The galaxy age, from the onset of  the star formation to $z$=0, is 13\,Gyr. 
When weighted by B-band luminosity the recovered age is  of 7 Gyr. Its
total mass, including the dark matter,  is 5.79$\times$10$^{10}$ M$_\odot$,  that of stars
3.75 $\times$10$^{10}$ M$_\odot$ within D$_{25}$.
The simulation  predicts 6.7\,Gyr between the maximum value of its star formation 
rate, i.e. the brightest point of the color magnitude diagram, and the following quenching.  
From z$\sim$1, when  the  stellar mass is prevailing over the mass 
of the dark matter inside  a radius  R$_{25}$,  the galaxy 
assembles about 54\% of its present stellar mass, and only 15\% from $z=0.5$ 
to $z=0$.

NGC~1533,  classified (RL)SB0$^0$ \citep[][]{Comeron14},
 has a  prominent UV-bright incomplete ring structure (Figure~\ref{fig:1b}). 
It is located in a clump of the Dorado group ($\rho=0.88$
galaxies Mpc$^{-3}$, \citet{Tully88}), in a large cloud of HI 
(7.4$\times10^{9}$ M$_\odot$ see Table~\ref{tbl-1}).

\citet{Mazzei14b} provided details on the evolution of NGC 1533, 
from the simulation accounting for its global properties. The
simulation succeeded in reproducing its  absolute magnitude,  
SED, morphology including the ring (their Fig. 6),  
with a merger between 
halos with mass ratio 2:1 and perpendicular spins.  
The predicted galaxy age  is 13.7\,Gyr, 
that estimated from weighting  the age of its stellar populations by B-band  luminosity
is 6\,Gyr,  younger than that of NGC 1426.
Using line-strength indices, \citet{Annibali07} calculated
a luminosity weighted age of 11.9$\pm$6.9 Gyr  with the large uncertainty due to the
presence of emission lines  which perturb the measure of H$\beta$. 
However, the far and near ultraviolet emissions revealed by GALEX \citep{Marino11c} 
and by our {\it Swift}-{\tt UVOT}  images suggest quite recent SF events in 
the outer ring of this galaxy.

Summarizing, the local ($z=0$) global properties of both NGC~1426 and NGC~1533,
 the first without, the second with peculiar features in the ultraviolet respectively, 
are explained in terms of a major merger event. 
In these cases, together with multi-wavelength photometric data,
high resolution velocity fields, derived by our 2D wide field Fabry-Perot observations, 
are available to constrain our simulations.

\subsection{Amidst true and `false'  galaxy pairs in groups}
\label{s:4.2}

Since the very beginning of extragalactic astronomy, about a century ago, pairs of galaxies 
fascinated  astronomers \citep[see e.g.][]{Lundmark1927} who
accumulated evidence  indicating that gravitational interaction 
affects the evolution of pair members \citep{Holmberg1937,Struck2011}.
Catalogues of nearby pairs prospered for a long time
 \citep[e.g.][]{Karachentsev1972,Turner1976,RR95,Soares1995}, however,
identifying physical pairs, i.e. gravitationally bound objects, 
has been non-trivial even after redshifts have become available. 
\citet{Karachentsev1989} 
calculated that in his catalogue 32\% are ``false" double systems, i.e.  
the two galaxies  have similar redshifts but are physically un-bound. 

%% Figure 6
%
\begin{figure}[tb]
\includegraphics[width=\columnwidth]{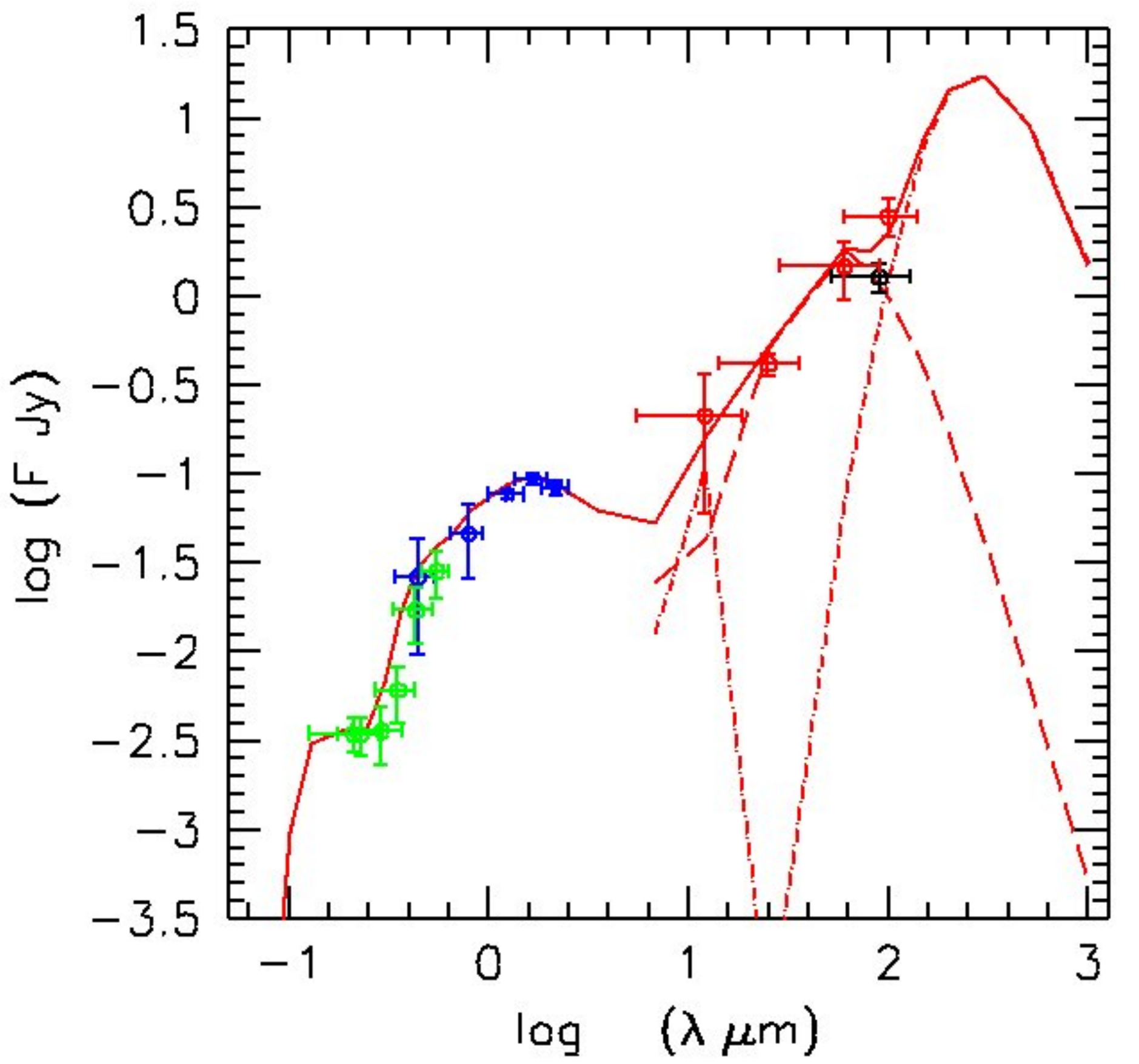}
\caption{ Dots of different colors are total fluxes of 
NGC~454 total system, the (red) solid line is the SED 
predicted by the SPH-CPI simulation best fitting the 
observations. Dashed and dot dashed lines 
represent the contribution  of the {\it warm} and
{\it cold} dust component, respectively, to the simulated SED
\citep[][]{Plana17} 
} %% no full stop at the end of caption
\label{fig:5}
\end{figure}
%%%%%%%%%%%%%

In the context of the evolution of galaxies in nearby groups, 
we have studied the  particularly intriguing case of the NGC~3447/3447A. 
This system is located in a Local Group Analog, LGG~225,
i.e. a group that, as the Local Group of Galaxies, is mostly populated
by late-type galaxies  with few of them dominant in the group
\citep[see e.g.][]{Marino2010}. The
NGC~3447/3447A system morphology, shown in the left panel of Figure~\ref{fig:4},
has been interpreted as a pair distorted by an on-going interaction. In our recent study  
\citep{Mazzei17}, we investigated  its optical and ultraviolet surface photometry
revealing that NGC~3347 has an extended disk that includes NGC~3447A.
The 2D velocity  field and velocity dispersion map we obtained 
using the Fabry-Perot PUMA@2.2m in San Pedro Martir (Mexico) suggest 
 that neither NGC~3447 nor NGC~3447A have
a coherent rotation velocity, but rather that there is a 
small velocity gradient between the two. These multi-wavelength data allowed us  to
constrain our large  grid of  SPH-CPI  simulations.  In the right panel Figure~\ref{fig:4} 
we show the far-ultraviolet image of NGC~3447/3447A derived from the
snapshot of the simulation provides an excellent match to the 
ultraviolet morphology and brightness, in addition to the best fit of the system 
SED, the B-band magnitude, and  the best match of the velocity field.
According to this simulation, the peculiar morphology of the system
can  be interpreted as a consequence 
 of the disk instability driven by the halo instability itself, enhanced by the 
on-going encounter  with a distant ($\sim1$ degree) companion,  0.5 mag fainter.
The simulation suggests that NGC~3447/NGC~3447A is a single object. 
The predicted age of the system is 12\,Gyr, while its stellar populations,  
weighted by B-band luminosity, are 1.3\,Gyr old. NGC 3347A/NGC 3347 
represents a new class of false pairs, that is, galaxies which appear as pairs but are a single distorted galaxy. 
\citet{Karachentsev1972, Karachentsev1989} warned about the potential danger of an 
underestimation of the role of false pairs since they gave "anomalously high
values of average orbital mass for binary galaxies". 
The problem reverberates to new large pair surveys. 
The evolution of this  dark matter-dominated system \citep{Mazzei01, Mazzei17}, 
which results from an encounter with a still far away companion  
giving rise to a very common intermediate luminosity galaxy,  
is a building block in galaxy assembly.\\

We also analysed  the NGC 454 system 
 \citep[RR23 in the catalogue of Southern isolated pairs of][]{RR95} 
 which unlike  NGC 3447/3447A system, is a physical pair.  
 This is composed of  a distorted early-type galaxy and a 
 completely damaged late-type galaxy so that it is considered an 
 example of on-going mixed merger. We
exploited the {\it Swift}-{\tt UVOT} archival optical and near ultraviolet
photometric data and our 2D kinematical information 
obtained with the SAM+FP at the SOAR 4.1m 
telescope at Cerro Pach\'on, Chile. This 
high spatial resolution Fabry-Perot provided us with
the H$\alpha$ intensity distribution, 
the 2D velocity and velocity dispersion fields \citep{Plana17}.  
The two members in the NGC 454 
system, nearly superposed in projection, result to have  a virtually null 
velocity separation. The velocity field of the late-type member 
is characterised by a small velocity gradient without any rotation pattern. 

We used {\tt UVOT} observations to  the SED of the system
 towards the near ultraviolet (see Fig.\ref{fig:5}) and to demonstrate 
 quantitatively that NGC~454 has a disk: S\'ersic indices run from $n_{W2}=1.09$ to 
 $n_{V}=1.79$.  The early-type member of the NGC 454 system is a {\it bona fide} 
 S0.  The global properties of this system, that is its total B-band absolute magnitude, 
 multi-wavelength SED, and morphology, are well matched by an on-going major merger 
between halos with a mass ratio  1:1 and perpendicular spin. In the current phase of the
encounter, the far-IR SED
accounts for a B-band attenuation of 0.85 mag so that the absolute
B-band magnitude of the best-fit snapshot, shown in Figure~\ref{fig:5},  
is -21.0 mag  to be compared with -20.64$\pm$0.41 mag.
About 33\% of the far infrared emission is due to a warm dust component, heated by
the ultraviolet radiation of H\,{\textsc{II}}  regions, and a  cold component heated
by the general radiation field, both including PAH molecules as
described in \citet{Mazzei1992, Mazzei1994} and \citet{MazzeideZotti1994}. 
The shape of the far infrared SED suggests the 
presence of  a large amount of dust:  the ratio between the  far
infrared luminosity  and the observed luminosity,
in the ultraviolet to near infrared spectral range, is 2.5.
 
Our SPH-CPI simulation predicts that this system is 12.4 Gyr old 
and that  the merger will be completed  within 0.2 Gyr 
\citep{Plana17}.

As expected, both exemples point to the conclusion that 
encounters are the driver of the galaxy morphological and physical
evolution in groups.

\section{Summary and conclusions}
\label{C}

We have investigated  galaxy groups at the two extremes of their 
evolution: i.e groups still non virialized, likely at the 
beginning of their evolution/assembling (as USGC U268), and
groups in  an  advanced phase as the NGC~5846 group.

We have analysed  the ultraviolet structure of their early-type galaxy members 
to explore the presence of recent star formation episodes. 
Furthermore, we have used the wealth of multi-wavelength information
and, when available, the 2D kinematics to constrain our large grid of SPH-CPI 
simulations and shed light on  the evolution of these galaxies.
The main results of  our works can be summarised 
 as follows : 

\begin{itemize}
\item {Bright early-type galaxies located in the Red Sequence 
of galaxy groups of different richness and  dynamical evolutionary phase 
are not passively evolving \citep{Yi2011, Marino16}; several of these galaxies
host recent star formation both in their nuclei and in their 
outskirts, often showing ring/arm-like structures.}

\item{A synoptic  view of the behaviour of the S\'ersic indices from
ultraviolet to optical and near-infrared bands of 11 early-type galaxies unveils
the presence of an underlying ultraviolet disk highlightening the role of 
dissipative mechanisms along the whole evolution of these galaxies \citep{Rampazzo17}.
Larger samples are needed  to draw firm conclusions.  }

\item {Our SPH-CPI simulations are able to reproduce 
the multi-wavelength properties  of early-type galaxies, i.e. their absolute 
magnitudes, SEDs, morphology including ultraviolet bright rings. They
 predict that up to 30\% of their stellar mass is assembled  
 during the galaxy transition from the  
Blue Cloud to the Red Sequence, i.e. from $z=1$ to $z=0$.
This transition takes several Gyrs  depending on galaxy luminosity, about
3-5\,Gyr  for the brightest ones.  Fainter galaxies may experience rejuvenation 
episodes more frequently than the brightest and massive ones \citep{Mazzei14a}.} 

\item {These simulations indicate that 
 Es in LGG~225 are  at least 1 Gyr younger than those in 
USGC~376 \citep{Mazzei14a}.
NGC~3457 is the result of a galaxy encounter and 
NGC 3522 of a major merger, while all the Es in USGC~376 derive 
from a merger.}

\item{At odds with rich and dense cluster environments, 
where external gas removal 
(e.g. ram-pressure stripping) are the most credited mechanisms
\citep[see][and references therein]{Poggianti2017}, our  SPH-CPI 
simulations suggest that star formation quenching in low density
environment is intrinsic to the galaxy evolution. }

\item{The study of pairs with our multi-wavelength
approach constrains  SPH-CPI simulations providing unexpected scenarios,
like the importance of disk instabilities during interactions leading to
{\it false pairs}, as in the case of NGC 3447/3447A \citep{Mazzei17}.}
\end{itemize}

\section{Future Directions}

We are enlarging our sample of early-type galaxies with ultraviolet,
mainly using the near ultraviolet {\it Swift}-{\tt UVOT} archived
data, optical and near infrared 
data, to perform multi-wavelength structural
analysis  with the purpose of testing our result 
that the S\'ersic index decreases at the ultraviolet wavelengths. 
We remark that both a PSF comparable with optical data
and large field of view are necessary ingredients
for an accurate ultraviolet surface photometry of galaxies in the nearby
Universe. Recent deep optical imaging shows
how  the investigation of outskirts may tell us about the galaxy
evolution \citep[see e.g.][]{Abraham2016,Duc2016}. 
Ultraviolet imaging has proven, even with very small telescopes, to 
 be indispensable for detecting and for interpreting some galaxy components 
 (youngest stellar populations, lowest star formation rates), in regimes 
 where optical and near infrared images [from the largest telescopes] had failed to 
 detect such components. This indicates that more powerful ultraviolet
  instruments (size, resolution and field of view) are obviously bound 
  to yield fundamental advances and to possibly open new paths. \\

The {\tt UVIT} \citep{Tandon2017} ultraviolet sensitive cameras with a field of view 
of 28\arcmin\ diameter and a PSF of about 1\arcsec\  is
a great improvement over the roughly 3\arcsec\ PSF  of  {\it Swift}-{\tt UVOT}. 
The field of view of the {\it Cosmological Advanced Survey Telescope for 
Optical and Ultraviolet Research} ({\tt CASTOR}) \citep{Cote2012} of 0.473$^\circ\times0.473^\circ$, with
three simultaneous near ultraviolet, U and $g$ filters and a HST-like spatial
resolution, would have a deep impact  on nearby galaxy studies, allowing
surface photometric analysis of both  more compact objects 
and  of galaxy associations.
Progress in the field of galaxy evolution would greatly benefit 
from optical designs that combine
both wide field of view and high resolution  imaging cameras for the 
next generation  of ultraviolet telescopes.

%% Math 
%
%\begin{eqnarray}%\label{eqn:?}
%\\ \nonumber
%\end{eqnarray}
%
%\begin{equation}%\label{eqn:?}
%\end{equation}

%% Table (two-column)
%
% \begin{table*}%%[tb]
% \small
% \caption{Caption} %% no full stop at the end of caption
%  \label{tbl:?}
% \begin{tabular}{}
% \tableline  %% rule at top
% \tablenotemark{a} 
% <entries>
% \tableline %% rule at bottom
% \end{tabular}
%
%% Any table notes must follow the \end{tabular} command. 
% \tablenotetext{a}{}
% \tablecomments{}
% \end{table*}

%% Table (one-colum)
%
% \begin{table}
% \caption{} %% no full stop at the end of caption
% \label{tbl:?}
% \begin{tabular}{}
% \tableline  %% rule at top
% \tablenotemark{a} 
% <entries>
% \tableline %% rule at bottom
% \end{tabular}
% \end{table}

%% Deluxe tabel (refer to AASTeX documentation)
%
% \begin{deluxetable}{ccrrrrrrrrcrl}
% \tabletypesize{\scriptsize}
% \rotate
% \tablecaption{}% %% no full stop at the end of caption
% \label{tbl:?} 
% \tablewidth{0pt}
% \tablehead{\colhead{}}
% \startdata
% <entries>
% \enddata
%
%% Text for table notes should follow after the \enddata but before
%% the \end{deluxetable}. Make sure there is at least one \tablenotemark
%% in the table for each \tablenotetext.
% \tablecomments{}
% \tablenotetext{a}{}
% \tablenotetext{b}{}
% \end{deluxetable}

%% Figure 
%
% \begin{figure}%[tb]
% \includegraphics{}
% \includegraphics[width=\columnwidth]{}
% \caption{} %% no full stop at the end of caption
% \label{fig:?}
% \end{figure}

%% Acknowledgements
%
 \acknowledgments
P. Mazzei and R. Rampazzo acknowledge partial support
 from INAF through grants PRIN-2014-14 `Star formation 
and evolution in galactic nuclei" (PI Michela Mapelli). 
This research has made use of the NASA/IPAC Extra galactic Database ({\tt NED}) 
which is operated by the Jet Propulsion Laboratory, California Institute of 
Technology, under contract with the National Aeronautics and 
Space Administration. Part of this work is based on archival data, 
software or online services provided by the Space Science Data Center - ASI.
%% References
%% Please cite all reference entries in the article text using \cite or
%% equivalent command. 

%%%  Using BibTeX  (Name-Year style)
%
% \bibliographystyle{spr-mp-nameyear-cnd}  %% BibTeX style
% \bibliography{<bib data>}                %% BibTeX data

%% Non-BibTeX  (Name-Year style)
%

\end{document}